\documentclass[a4paper,11pt]{article}

\usepackage[top=0.8in, bottom=0.7in, left=0.9in, right=0.9in]{geometry}

\usepackage{titlesec}

\usepackage{multicol}
\usepackage{floatrow}
\newfloatcommand{capbtabbox}{table}[][\FBwidth]
\usepackage{fourier}
\usepackage{color}
\usepackage{graphicx}
\usepackage{url}
\usepackage[affil-it]{authblk}
\usepackage{amsmath}
\usepackage{wrapfig}
\usepackage{booktabs}
\usepackage{tabularx}
\usepackage{caption}
\usepackage{subcaption}
\usepackage{comment}
\usepackage{academicons}
\usepackage{hyperref}

\newbox{\bigpicturebox}

\usepackage[T1]{fontenc}
\usepackage{times}

\newcolumntype{P}[1]{>{\centering\arraybackslash}p{#1}}
\DeclareCaptionLabelFormat{andtable}{#1~#2  \&  \tablename~\thetable}

\titlespacing*{\section}{0pt}{1.6ex plus 1ex minus .2ex}{1.6ex plus .2ex}
\titlespacing*{\paragraph}{0pt}{1.6ex plus 1ex minus .2ex}{1.6ex plus .2ex}

\begin{document}

\title{Evaluate Fine-tuning Strategies for Fetal Head Ultrasound Image Segmentation with U-Net}

\author[1]{Fangyijie Wang}
\author[2]{Gu\'enol\'e Silvestre}
\author[1]{Kathleen M. Curran}

\affil[1]{School of Medicine, University College Dublin, Dublin, Ireland}
\affil{fangyijie.wang@ucdconnect.ie}
\affil[2]{School of Computer Science, University College Dublin, Dublin, Ireland}
\date{}
\maketitle
\thispagestyle{empty}

\begin{abstract}
Fetal head segmentation is a crucial step in measuring the fetal head circumference (HC) during gestation, an important biometric in obstetrics for monitoring fetal growth. However, manual biometry generation is time-consuming and results in inconsistent accuracy. To address this issue, convolutional neural network (CNN) models have been utilized to improve the efficiency of medical biometry. But training a CNN network from scratch is a challenging task, we proposed a Transfer Learning (TL) method. Our approach involves fine-tuning (FT) a U-Net network with a lightweight MobileNet as the encoder to perform segmentation on a set of fetal head ultrasound (US) images with limited effort. This method addresses the challenges associated with training a CNN network from scratch. It suggests that our proposed FT strategy yields segmentation performance that is comparable when trained with a reduced number of parameters by 85.8\%. And our proposed FT strategy outperforms other strategies with smaller trainable parameter sizes below 4.4 million. Thus, we contend that it can serve as a dependable FT approach for reducing the size of models in medical image analysis. Our key findings highlight the importance of the balance between model performance and size in developing Artificial Intelligence (AI) applications by TL methods. Code is available at \href{https://github.com/13204942/FT_Methods_for_Fetal_Head_Segmentation}{GitHub}.
\end{abstract}

\section{Introduction}
Training a deep CNN from scratch can prove to be a formidable undertaking, particularly in medical applications that are often constrained by limited annotated data and require a substantial time investment. However, Transfer Learning (TL) can help alleviate these challenges. TL is a technique in which a network learns from a large dataset and then applies that knowledge to another application, typically a smaller dataset. This approach can be especially advantageous in medical applications where annotated data is scarce, as it permits the utilization of pre-trained models to enhance performance on smaller datasets. TL approaches entail the adoption of pre-trained models and fine tuning (FT).

\begin{sloppypar}
In this study, we conducted a segmentation task on fetal head US images using deep neural networks with various FT strategies. The dataset HC18 comprises of 1334 ultrasound images obtained from 551 pregnant women and is publicly available \cite{Heuvel:2018}. To perform semantic segmentation on the HC18 fetal head US images, we performed the FT of the U-Net \cite{Ronneberger:2015} network, with a pre-trained MobileNet \cite{Howard:2018} as its backbone. In order to develop a lightweight model using FT techniques, this research work considered a comparison of model sizes for various pre-trained CNN models. Furthermore, we investigated the impact of FT on different decoder layers for fetal head segmentation. In terms of segmentation outcomes on tests, the results were competitive in comparison to the state-of-the-art (SOTA) results, 97\% ($\pm$ 0.3\%) achieved by \cite{Amiri:2020} with FT the encoder. Our research is of significance when analyzing the trade-off between performance and model size in the development of mobile AI applications.
\end{sloppypar}

The main contributions of this paper are as follows: (1) We analyzed eight different fine-tuning strategies on a U-Net network that used a MobileNet V2 encoder to predict segmentation masks from a fetal head ultrasound dataset. (2) We achieved SOTA accuracy on the HC18 Grand Challenge by providing a pre-trained U-Net model that had only 4.4 million trainable parameters. (3) Our experiments showed that unfreezing the decoder of a pre-trained U-Net network was the most effective fine-tuning strategy compared to the others we tested.

\section{Related Work}

\begin{sloppypar}

In recent years, DL techniques have been developed to achieve high precision outcomes in semantic segmentation tasks. Ronneberger et al.~\cite{Ronneberger:2015} proposed the U-Net architecture to perform biomedical image segmentation tasks with annotated samples more efficiently. In 2019, Howard~et al.~\cite{Howard:2018} constructed MobileNet V2 for semantic segmentation by making use of lightweight depth-wise separable convolutions to filter features. Therefore, it has a lower computational cost, less memory, and consumes less power. As a result, MobileNet V2 is a low-cost, efficient deep neural network suitable for mobile and embedded vision applications.

In terms of US image segmentation tasks, \cite{Amiri:2020} employs TL techniques to overcome limited and costly data issues in DL for medical applications. The authors investigate the impact of FT various layers of a pre-trained U-Net and assess their performance in fetal US image segmentation tasks on the HC18 US dataset. Their FT strategies consist of three schemes, FT shallow, deep layers, and the entire network. Across all US datasets analyzed in their work, FT the entire pre-trained U-Net yielded better results than training from scratch. \cite{cheng:2021} utilizes cross-domain TL with U-Net architecture for precise and fast image segmentation. The cross-domain TL techniques are utilized in \cite{Monkam:2023} for the purpose of fetal head segmentation on HC18. The researchers have proposed a speedy and efficient method to produce a considerable number of annotated US images, based on a limited number of manually annotated biometrics. Besides cross-domain TL techniques, Alzubaidi et al. \cite{Alzubaidi:2022} demonstrated an ensemble TL technique with a segmentation model that includes eight CNN models. This technique is evaluated on the US dataset HC18 by achieving 98.53\% mIoU. However, the ensemble TL model has 28.12 million trainable parameters, which is 7 times more than the best model we proposed with 4.4 million trainable parameters. \cite{kim:2022} provides an overview study of TL methods on medical image classification. They demonstrated the efficacy of TL. The authors suggest that utilizing CNN models as feature extractors can save computational costs. Inspired by the investigation from \cite{kim:2022}, we think similar FT methods can be utilized in medical image segmentation.

Our proposed FT strategy achieved competitive head segmentation results on HC18 with fewer trainable parameters and training epochs compared to existing SOTA methods, see Figure \ref{fig:sota_comparison}. The U-Net is a strong CNN architecture widely applied in medical image analysis. The most notable segmentation outcomes on the present HC18 leaderboard were obtained by leveraging U-Net and its expansion networks. Hence, we utilize U-Net architecture to construct a CNN model and evaluate our FT strategies.

\end{sloppypar}

\section{Methodology}

\paragraph{Data Preparation:} 
The HC18 dataset comprises a two-dimensional (2D) US image collection that has been split into 999 images for training purposes and 335 images for testing. All HC18 images are standard planes that are suitable for measuring fetal HC. Each of these images is of dimensions 800 by 540 pixels. Because these 999 images were annotated with biometrics by experienced medical experts, they were selected as the experimental dataset, whereby 799 images and 200 images were assigned for training and testing, respectively. 999 images were resized to 512 $\times$ 512 pixels. In this study, we used the standard data-augmentation techniques: rotation by an angle from $[-25^{\circ}, 25^{\circ}]$, horizontal flipping, vertical flipping, and pixel normalization.

\paragraph{Model Design:} 
In our work, based on Ronneberger's work \cite{Ronneberger:2015}, we built a U-Net baseline model with 4 encoder layers, 4 decoder layers and 1 bottleneck. The model has input features [64, 128, 256, 512]. We apply a MobileNet V2 model to the U-Net's encoder part. The MobileNet V2 model was pre-trained on dataset ImageNet.

\begin{figure}
     \centering

    \sbox{\bigpicturebox}{%
    \begin{subfigure}[b]{.65\textwidth}
        \scalebox{1}[1]{\includegraphics[width=\textwidth]{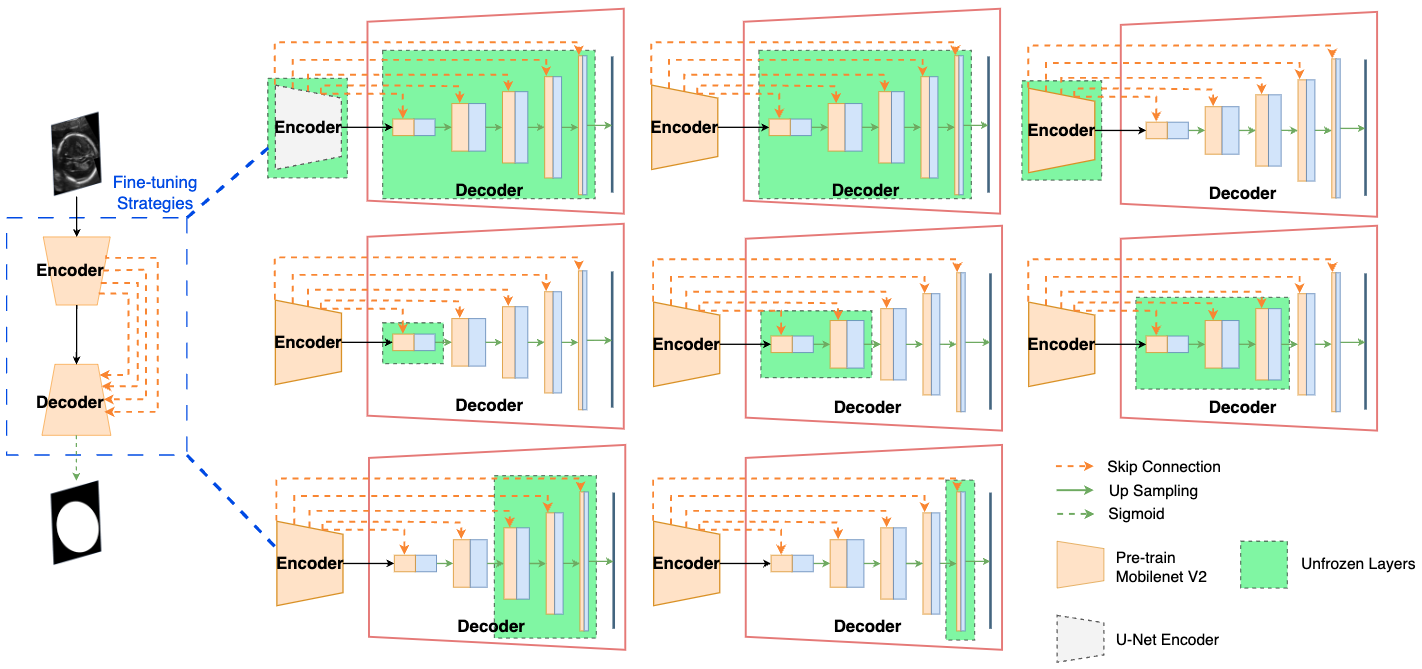}}%
        \vspace{-0.3em}
        \caption{}
        \label{fig:ft_methods}
    \end{subfigure}
    }
    \usebox{\bigpicturebox}
    \hfill
    \hspace{-1.4em}
    \begin{minipage}[b][\ht\bigpicturebox][s]{.34\textwidth}
    \begin{subfigure}{1\textwidth}
        \includegraphics[width=\textwidth]{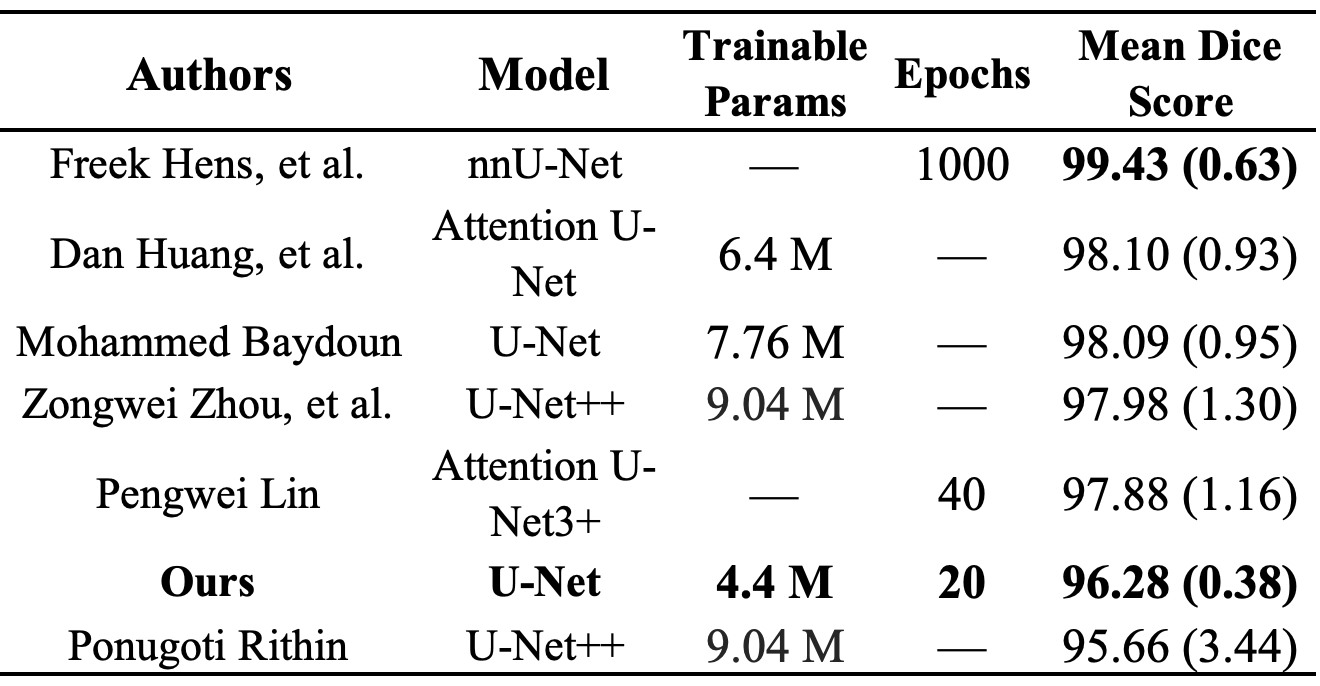}
        \vspace{-1.7em}
        \caption{}
        \label{fig:sota_comparison}
    \end{subfigure}
    \vfill
    \begin{subfigure}{1\textwidth}
        \includegraphics[width=\textwidth]{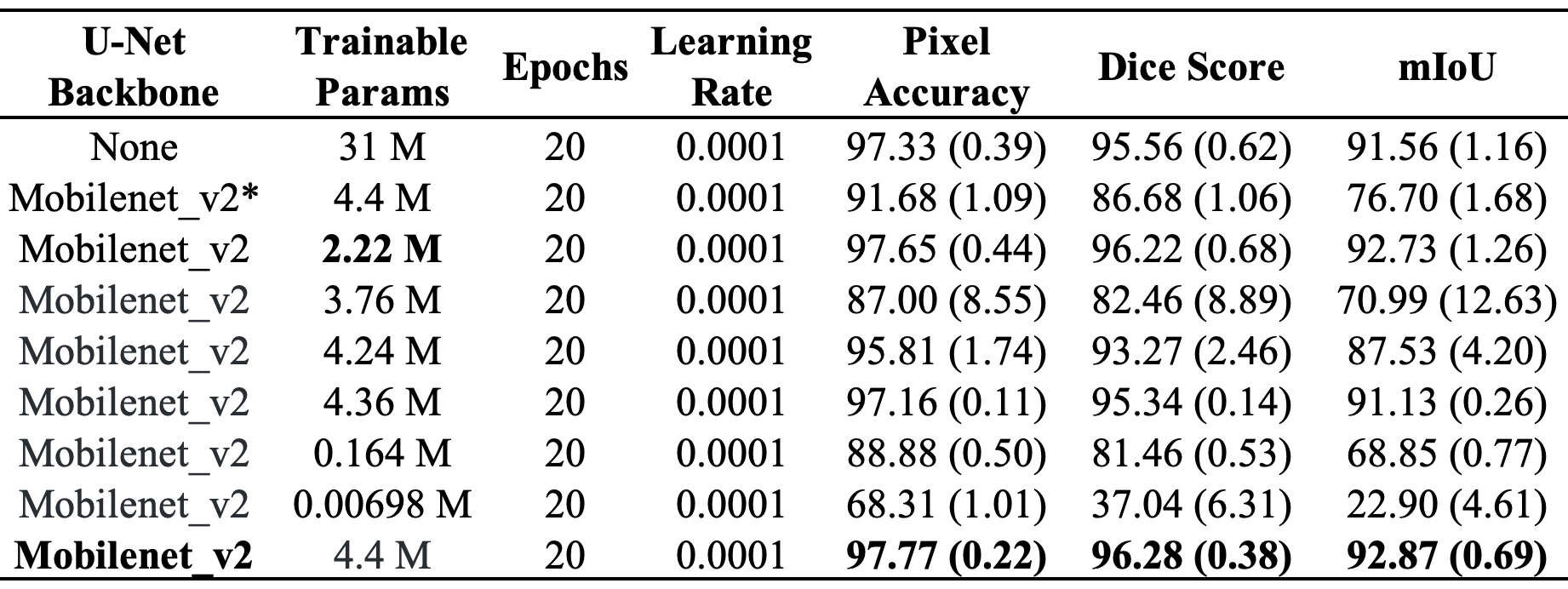}
        \vspace{-1.5em}
        \caption{}\
        \label{fig:numeric_results}
    \end{subfigure}
    \end{minipage}
    \vspace{0.4em}
    \caption{(a) The first row shows three fine-tuning strategies: U-Net baseline, 0 to 4 layers remain unfrozen within the decoder, and the encoder remains unfrozen. The second row shows three fine-tuning strategies: 0 layer remains unfrozen within the decoder, 0 to 1 layers remain unfrozen within the decoder, and 0 to 2 layers remain unfrozen within the decoder. The last row shows two fine-tuning strategies: 2 to 4 layers remain unfrozen within the decoder, 4 layer remains unfrozen within the decoder. (b) Comparison of our methods with the SOTA methods. (c) Comparison of Pixel Accuracy, Dice Score, and mIoU on Test data set. Mobilenet\_v2\(\ast\) is the encoder with random weights.}
    \label{fig:two_graphs}
\end{figure}

\paragraph{Fine-tuning Strategies:} 
Our FT methods include a collection of seven distinct schemes, see Figure \ref{fig:ft_methods}. The baseline U-Net model has no pre-trained encoder and all layers unfrozen. The FT methods include training the entire decoder, the entire encoder, 0 layer within decoder, 0,1 layers within decoder, 0,1,2 layers within decoder, 2,3,4 layer within decoder, and 4 layer within decoder. In the baseline U-Net model, the encoder is not pre-trained and all layers remain unfrozen. The FT methods are comprised of a range of techniques, including training the entire decoder, the entire encoder, the layer 0 within the decoder, layers 0 to 1 within the decoder, layers 0 to 2 within the decoder, layers 2 to 4 within the decoder, and the layer 4 specifically within the decoder. In all experiments, the training and testing operations are executed four times repeatedly.

\paragraph{Training and Evaluation:}
We implemented all of our experiments using Pytorch. After comparing performance between different CNN architectures, we train a U-Net model on HC18 from scratch by using Segmentation Models \cite{Iakubovskii:2019}. We trained the U-Net model with 20 epochs from scratch. Each epoch took around 75 seconds. Also, we fine-tuned the pre-trained U-Net with MobileNet V2 encoder with 20 epochs. Each epoch took around 25 seconds. The training dataset and test dataset both have a batch size of 10. The Adam optimiser was used in training processes with a decaying learning rate of \(1e-4\). All training processes were performed on an NVIDIA Tesla T4 graphics card. The typical metrics applied to evaluate the performance of segmentation models are Pixel Accuracy (PA), Dice coefficient, and Mean Intersection over Union (IoU). Mean IoU is defined as the average IoU over all classes \(K\).

\section{Experimental Results}

Figure \ref{fig:numeric_results} summarises the segmentation metrics achieved through the implementation of various FT strategies on the HC18 test set, 200 fetal US images. The act of unfreezing the entire decoder within the pre-trained U-Net model has contributed to the generation of more accurate predictions on segmentation masks when compared to both the U-Net baseline model and other FT strategies. Our proposed FT strategy improved PA, Dice score, and mIoU by 0.45\%, 0.75\%, and 1.4\% respectively when compared to training our U-Net baseline from scratch. Furthermore, the size of trainable parameters has been reduced by 85.8\%. Despite the fact that the size of trainable parameters for other FT strategies is smaller than 4.4 million, our proposed FT strategy outperformed their evaluation results. In comparison to Amiri's methods, our proposed FT strategy has also yielded a 1.24\% increase in their results (95.1\%) \cite{Amiri:2020} in terms of Dice score. Another FT strategy involving training U-Net pre-trained encoder only has also shown competitive results with a 96.22\% Dice score.

\section{Conclusion}

We presented a FT strategy for a pre-trained U-Net that enables accurate fetal head segmentation in US images while utilizing only 4.4 million parameters. To evaluate the effectiveness of various fine-tuning approaches, we conducted experiments on the HC18 Grand Challenge dataset. Our findings suggest that utilizing a pre-existing network enhances segmentation precision, whereas augmenting the amount of trainable parameters does not significantly impact accuracy. To reduce model size and the number of trainable parameters, we used the MobileNet V2 model as the encoder in our U-Net. Our fine-tuned model has significantly reduced 85.8\% trainable parameters in comparison to training an initialized U-Net. Our research suggests that the ideal approach for FT is to adjust the decoder's 0, 1, 2, 3, 4 layers of the pre-trained U-Net based on our experiments. This methodology yielded a PA of 97.77\%, a Dice coefficient of 96.28\%, and a mIoU of 92.87\% on the HC18 test dataset. Alternatively, FT the U-Net pre-trained encoder only is another TL method producing competitive results potentially. Our findings propose that adjusting the decoder of the U-Net might serve as an efficient approach for FT small models in US image analysis.

\section{Future Work}

Future research may be conducted in order to reduce noise on US images by introducing image processing methods. And we will further investigate the resilience of the model that has been trained by TL techniques. Furthermore, we intend to investigate alternative pre-trained models in order to achieve an optimized model that is smaller in size. 

\section*{Acknowledgments}

This publication has emanated from research conducted with the financial support of Science Foundation Ireland under Grant number 18/CRT/6183. For the purpose of Open Access, the author has applied a CC BY public copyright licence to any Author Accepted Manuscript version arising from this submission.


\bibliographystyle{apalike}
\bibliography{imvip}

\end{document}